# Orientation-dependent high harmonic generation in ZnO

Tzveta Apostolova[1,2], Balint Kiss[3], Deborata Rajak[3], Katalin Pirisi[3], Rajaram Shrestha[3], Levente Abrok [3], Boyan Obreshkov[1], Jean Claude Kieffer[4], Dimitar Velkov[5]

[1]Institute for Nuclear Research and Nuclear Energy, Bulgarian Academy of Sciences
[2] Institute for Advanced Physical Studies, New Bulgarian University
[3]ELI-ALPS, ELI-HU Ltd
[4]ELI-ERIC
[5]Sofia University

*ABSTRACT*

*We theoretically and experimentally investigate orientation-dependent high-harmonic generation (HHG) in zinc oxide subjected to intense femtosecond mid-infrared (MIR) laser pulses. In agreement with past measurements from literature, we observe non-perturbative harmonic spectra with harmonics extending beyond the material band gap. The spectra depend sensitively on the orientation of the crystal with respect to the laser polarization direction. To benchmark the measurements, we performed detailed theoretical calculation of the orientation-dependent harmonic yields. The theory predicts a three-fold angular modulation of the intensity of odd-order harmonics in the (0001)-plane, which can be deduced from the symmetry of the hexagonal lattice alone, in contrast a four-fold symmetry is observed in the experiment. The theory points out the essential role of the anisotropy of the multiple band structure affecting the dynamics of electron-hole pair excitation and solid-state HHG.*

## Introduction

High-harmonic generation (HHG) serves as a primary technique for generating coherent extreme-ultraviolet radiation and attosecond pulses. While initially demonstrated in gases, HHG in solids offers a pathway toward compact, bright, and tabletop EUV sources, providing new opportunities to probe ultrafast electron dynamics in crystalline and amorphous materials [1], to perform molecular tomography [2,3], and to carry out high-resolution spectroscopy [4]. The use of mid-infrared (MIR) driving pulses is particularly advantageous since the ponderomotive energy of electrons scales with the square of the wavelength [5], allowing electrons to reach high kinetic energies even at moderate field strengths [6-8]. Modern laser systems deliver pulse energies in the tens of millijoules [9], high repetition rates [10], and durations approaching a few optical cycles [11], creating ideal conditions for exploring strong-field phenomena in solids.

Zinc oxide (ZnO) has emerged as a key material for studying HHG in solids. Experiments have shown that the crystallographic orientation relative to the laser polarization strongly modulates the harmonic spectrum [12–14]. Both even and odd harmonics are typically observed, but along symmetry axes where reflection or inversion symmetry is preserved, even harmonics are suppressed, while breaking these symmetries restores their presence [15]. Two-color driving schemes, combining a strong MIR fundamental with a weaker second-harmonic field, further enable selective control of even and odd harmonics through temporal symmetry breaking [16,17].

Polarization- and orientation-resolved HHG thus provides a sensitive probe of the electronic structure, symmetry properties, and interband dynamics of solids [16-18]. For instance, momentum-resolved HHG measurements in ZnO have been used to reconstruct the bandgap along the Γ–A direction [16], while polarization-resolved studies reveal conduction-band dispersion [19-23] and Berry curvature effects [24].

The present work systematically investigates how the harmonic yield depends on drive pulse parameters - fluence, polarization and wavelength. Experimental results are complemented by theoretical simulations of electron–hole pair excitation and HHG under intense femtosecond MIR pulses, providing a detailed understanding of the mechanisms governing high-harmonic emission in solids.

**Experimental study of high harmonic generation in ZnO**

The experiments were performed using the mid-infrared laser system at the ELI-ALPS Research Facility, which delivers pulses centered at 3.2 μm with a repetition rate of 100 kHz. The output of the optical parametric chirped pulse amplification (OPCPA) system [25] provides pulse energies of approximately 100 μJ, which are compressed to 45 fs (~4 optical cycles) through nonlinear spectral broadening in a 2 mm thick yttrium aluminum garnet (YAG) crystal and a 1 mm thick Si plate, followed by a $BaF_2$ wedge pair. The carrier-envelope phase (CEP) of the pulses is passively stable and actively controlled using an acousto-optical phase modulator (Fastlite, Dazzler) with feedback from an f–2f interferometer combined with a CEP measurement unit (Fastlite, Fringezz).

The central wavelength of 3.2 μm enables access to strong-field scaling laws, including the quadratic wavelength dependence of the harmonic cutoff in high-order harmonic generation (HHG). The high repetition rate of 100 kHz allows rapid data acquisition with excellent statistics, while the stability of the system supports long-term measurements of processes with low cross-sections. The post-compressed few-cycle pulses with stabilized CEP are ideally suited for driving CEP-sensitive HHG in semiconductor crystals. The peak intensity and field strength were varied systematically in power scans, reaching values of a few TW/cm² while remaining below the reported damage threshold of ZnO [13].

Fig. 1 shows a schematic sketch of the experimental setup. In the setup the laser beam is focused using a protected gold-coated concave mirror with a focal length of 300 mm to a spot size of 150 μm by 136 μm onto the sample. The duration of the laser pulses was estimated be around 45 fs at FWHM. Samples are inserted into the focal plane of the laser beam using a Smaract vacuum compatible rotational wheel on an Owis linear stage (moveable in z direction). The high harmonics generated in the samples are recorded both in ambient air and in vacuum ($\sim 10^{-6}\ mBar$) using an Avantes Avaspec 3648 spectrometer and an Andor Newton CCD camera after being focused by two $MgF_2$ lenses with focal lengths 200 mm and 100 mm respectively. Additionally, the laser beam passes through combination of HWP and linear polarizer in order to adjust the power of the pulses and a through a combination of HWP and QWP to change the polarization state of the light before the interaction with the sample.

In Fig. 2 the wurtzite ZnO stable hexagonal structure is shown, in which each lattice cell consists of tetrahedral structures centered with an oxygen atom or zinc atom. The tetrahedral structure of ZnO is a consequence of the tetrahedral hybrid orbitals of oxygen and zinc atoms. The internal parameter u is defined as the length of the bond parallel to the c-axis (anion–cation bond length or the nearest-neighbor distance) divided by the c lattice parameter. The basal plane lattice parameter (the edge length of the basal plane hexagon) is denoted by a; the axial lattice parameter (unit cell height), perpendicular to the basal plane, is universally described by c.

In the first measurements, the sample of thickness 50 μm c-cut (0001) ZnO crystal was placed in the focal plane and irradiated in ambient air. Power scans were performed by varying the peak intensity (field strength) to establish the appearance of harmonic spectrum from the sample. The maximum peak intensity (field strength) used was approximately 1.2 TW/cm$^2$ and this intensity is below the reported damage threshold value of the crystalline solid. The position of the crystal and the half-wave-plate (HWP) were fixed for performing the power scans.

After these initial scans, the dependence of the high order harmonic spectrum on the driving laser polarization in respect to crystal orientation was measured, by rotating the HWP in the hexagononal plane of the sample. The generated high harmonic signal from each of the samples and for each measurement (data acquisition) was focused to the entrance slit of a UV-VIS spectrometer by an $f = 10$ cm $MgF_2$ lens. The Andor Newton CCD camera, which covers the EUV and XUV wavelength ranges, was used to record harmonics higher than the 13$^{th}$ harmonic order in vacuum.

# Experimental results on orientation dependence of high harmonic generation in ZnO

In Fig. 3 the angle-dependent high harmonic spectra from c- cut (plane 0001) ZnO sample of thickness 50 µm in transmission geometry consists of only odd harmonics and extends to 13$^{th}$ harmonic order. The laser polarization in the c-plane and is varied by rotating half wave plate placed in the laser beam path. In the right panel – only above band gap harmonics are shown. A four-fold symmetry is observed in the modulation of the intensity of the generated odd harmonics as a function of the linear polarization od the driving laser in the hexagonal plane where the angle of the HWP varies from $\theta = 0° - 180°$. The intensity of the generated harmonics exhibits peak intensity for angles - $\theta = 0°, 45°, 90°, 135°, 180°$. The used laser intensity $I = 1 TW/cm^2$.

Fig. 4 shows high harmonics spectrum of VUV harmonic orders generated in the c-cut ZnO sample of thickness 50 µm. The laser polarization in the c-plane and is varied by rotating half wave plate placed in the laser beam path. Odd harmonic orders of 13$^{th}$ harmonic order and above (up to 19$^{th}$) are exhibited. (Calibration needed). Four-fold symmetry of the modulation of the harmonic intensity is exhibited when varying the angle of polarization from $\theta = 0° - 360°$.

Fig. 5 displays the harmonic spectra from a- cut (plane 1120) ZnO sample of thickness 50 µm in transmission geometry. Since there is no inversion symmetry efficient even-order harmonics generation from the bulk occurs. The angle-dependent high harmonic spectra consists of both odd and even harmonics with below bandgap harmonics from 2$^{nd}$ to the 9$^{th}$ and the above bandgap harmonics extend to 13$^{th}$ harmonic order as registered by the UV-VIS spectrometer. The three panels display – left one – all generated odd and even harmonics up to 15$^{th}$ order, middle panel visualizes better harmonic orders ranging from 6$^{th}$ to 15$^{th}$ and the right panel displays only the above band gap harmonic orders. As for the c-cut sample of the same thickness, the intensity of the generated harmonics is strongly modulated by varying the linear polarization of the driving laser in the plane of the optical axis. A four-fold symmetry of intensity modulation is observed for odd harmonics for angles - $\theta = 0°, 45°, 90°, 135°, 180°$, and a two-fold symmetry in the intensity modulation of even harmonics is visible for angles $\theta = 45°, 135°$.

Fig. 6 shows high harmonics spectrum of VUV harmonic orders generated in the a-cut ZnO sample of thickness 50 µm. The laser polarization in the a-plane and is varied by rotating half wave plate placed in the laser beam path from $\theta = 0° - 360°$. Odd harmonic orders of 13$^{th}$ harmonic order and above (up to 19$^{th}$) are exhibited. Eight-fold symmetry of the modulation of the harmonic intensity is observed due to varying angle of polarization. Even harmonic orders of 12$^{th}$ harmonic order and above (up to 18$^{th}$) exhibit four-fold symmetry of intensity modulation as a function of the angle of the HWP in respect to crystallographic direction. As in ref. [11] even harmonics disappear for some angles of the HWP.

For a more complete comparison with the VUV spectra displayed in Fig. 6, we also present in Fig. 7 UV-VIS harmonic spectra generated by varying the angle of the HWP (laser linear polarization) in the optical plane of the a-cut ZnO sample of thickness 50 µm $\theta = 0° - 360°$. The three panels display – left one – all generated odd and even harmonics up to 15th order, middle panel visualizes better harmonic orders ranging from 6th to 15th and the right panel displays only the above band gap harmonic orders. Eight-fold symmetry of the modulation of the harmonic intensity is observed clearly due to varying angle of polarization for the odd harmonic orders and four-fold symmetry of the modulation of the harmonic intensity is observed clearly due to varying angle of polarization for the odd harmonic orders.

Fig. 8 shows the harmonic spectra from a- cut (plane 1120) ZnO sample of thickness 200 µm in transmission geometry. The angle-dependent high harmonic spectra consists of both odd and even harmonics with below bandgap harmonics from 2nd to the 9th and the above bandgap harmonics extend to 13th harmonic order as registered by the UV-VIS spectrometer. The three panels display – left one – all generated odd and even harmonics up to 15th order, middle panel visualizes better harmonic orders ranging from 6th to 15th and the right panel displays only the above band gap harmonic orders. A four-fold symmetry of the odd harmonics intensity modulation is observed for angles with nearly equal peak intensity and a two-fold symmetry of even harmonics intensity modulation is visible for angles $\theta = 45°, 135°$. As observed in ref. [11] the even harmonics are suppressed for some angles of the HWP in respect to the optical axis of the crystalline material. For this increased thickness, the harmonics exhibit a very distinct fine structure (satellite structure as for the 6th harmonic order – middle panel) and are much broader than the ones generated in the 50-µm thicknesses c-cut and a-cut samples. This broadness, structure and satellite structure of the harmonics can be attributed to propagation effects in transmission geometry. It can also be observed that the harmonic orders are slightly shifted towards higher values. For this thickness of the ZnO a-cut sample, the harmonic spectra display spectral shifts (both the below bandgap and above bandgap), which are possibly due to the shift in the instantaneous frequency of the diving laser field in transmission geometry. Their width increases significantly over the angle range of the HWP (much broader) and they display additional sharp peaked intensity structures within the defined harmonic orders over the HWP angles as well as satellite structures. The used laser intensity is $I = 1.2 TW/cm^2$.

### Theoretical results for orientation dependent HHG in bulk ZnO

The theoretical model of Bloch electrons interacting with pulsed laser electric fields is detailed in Ref. [27]. The ground state electronic structure of zinc oxide is described by the empirical pseudopotential method [28] and the band structure is presented in Fig. 9 together with the Brillouin zone of the wurtzite structure of ZnO. Below we discuss numerical results for photo-excitation and HHG in bulk ZnO subjected to laser pulse of mid-infrared wavelength with peak laser intensity 2 TW/cm$^2$ and pulse duration 90 fs (9-cycle pulse).

The laser polarization direction was varied in the (0001)-plane, and in the (11-20) plane of the crystal. On the (11-20) plane, the optical axis is along the c-axis, and the polarization angle $\theta$ of the laser is defined with respect to the optical axis. Figs.10 (a-c) show the energy-momentum dispersion of Bloch electrons in

ZnO for three different polarization angles. At the Γ point, there are two degenerate valence bands (v7,v8) derived from $2p_x$ and $2p_y$ orbitals of the oxygen atoms, the crystal field split-off $p_z$-like band (v6) is slightly below in energy from the valence band top. The direct band gap energy at the Brillouin zone center 3.3 eV, requires simultaneous absorption of at least nine laser photons to create an electron-hole pair. The three valence bands at the top disperse downwards as the magnitude of the crystal momentum $\vec{k}$ is varied along the optical axis ($\theta = 0°$), cf. Fig10 a. Besides the three valence bands at the top, there are two lower lying in energy valence bands (v4, v5), which are two-fold degenerate at the Γ point, and are separated from the top by 0.8 eV. For the MIR laser wavelength, the valence bands at the (v6, v7, v8) are strongly coupled to (v4, v5) bands by two-photon transition. The degeneracy among valence bands is lifted as the polarization angle is changed, when all relevant five coupled valence bands are revealed in Figs.10 (b, c).

In Fig.11 a-c, show the time-dependent probability for electron-hole pair excitation of the Gamma bands of electrons having conserved canonical crystal momentum $\vec{k} = (0,0,0)$. The laser polarization vector $\vec{e} = sin\theta \hat{b}_1 + cos\hat{b}_3$ is rotated in the (11-20) plane of the crystal, here $\hat{b}_1 = \left(\frac{\sqrt{3}}{2}, -\frac{1}{2}, 0\right)$ and $\hat{b}_3 = (0,0,1)$ is a pair of orthogonal unit vectors in that plane. The three valence bands at the top (v6,v7,v8) are coupled to the lowest in energy conduction band (c1), the pair of valence bands (v4,v5) are only indirectly coupled to the conduction band via their interactions with (v6,v7,v8) bands. When $\theta = 0$, see Fig.11 (a), there is only one relevant inter-band coupling allowed by selection rules, which connects (v6) to the (c1) band. The transition history of this inter-band transition was discussed in detail in Ref.[27] on the basis of two-band model. Here we reiterate some of the main features in the transient response of electrons: electron-hole pairs are born in short temporal bursts during each half-cycle of the drive pulse at the extrema of the oscillating laser electric field via non-adiabatic Landau-Zener transitions. Subsequently electrons and holes in a superposition of states $|\psi(t)\rangle = \sqrt{1-P}|v_6(t)\rangle + \sqrt{P}e^{iS(t)}|c(t)\rangle$ move in their respective bands and gain relative phase $S = \int_{t_0}^{t} dt' \Delta(\vec{k}(t'))$ upon propagation in the laser electric field, where $\vec{k}(t) = \vec{k} + \vec{A}(t)$ is the kinetic momentum, $\vec{A}(t)$ is the laser vector potential and $\Delta(\vec{k}(t))$ is the instantaneous band gap energy. Depending on the propagation phase, transition amplitudes from consecutive half-cycles of the drive laser may either reinforce each other and interfere constructively, or cancel each other out, such that the number of electron-hole pairs may either decrease or grow. In the considered example, there are few successive transitions near the pulse peak that more or less tend to cancel each other out, resulting in final transition probability not exceeding $10^{-4}$. In Fig.11 (b) ($\theta = 30°$), besides the dominant v6 → c1 interband transition, the promotion of electrons from the (v4, v5) bands into the conduction band becomes prominent. That is because once the valence bands at top (v6,v7,v8) are partially occupied due to tunnel transition into the conduction band, electrons from the lower valence bands (v4,v5) fill the holes in the upper three bands (v6,v7,v8) by resonant two-photon transition. Subsequently, these electrons emerge in the conduction band by tunnel transitions near the extrema of the drive laser pulse. This indirect process involving interaction among multiple valence bands occurs efficiently slightly after the pulse peak. In Fig.11(c) ($\theta = 90°$), the polarization vector lies in the intersection of the (11-20) with the (0001) plane, when the coupling of the (v6) band to the conduction band vanishes. However, the valence bands (v7, v8) are active, and by interacting with them via two-photon transition, the lower lying bands (v4, v5) undergo promotion into the conduction band.

The final probability of electron-hole pair excitation P (v→c) after conclusion of the pulse is shown as a function of the polarization angle $\theta$ in Fig12. For $\theta$ close to 0 or 180 degrees, the inter-band transition v6 → c1 gives dominant contribution to the photo-electron yield, note that the electron yield near $\theta = 180°$ is slightly higher than for $\theta = 0°$, because of the broken inversion symmetry of the ZnO crystal. As the angle reaches $\theta = 90°$, the electron yield decreases in magnitude, and the promotion of (v4, v5) bands becomes the dominant excitation mechanism.

Fig.13 shows the polarization angle dependence of the photo-induced inter-band current of electrons in the Gamma bands. During each half-cycle, the build-up of coherent superposition of states in the valence and the conduction bands creates a rapidly oscillating electric dipole, cf. Fig. 13 (a-c), which in turn emits photons in a range of energies near the bandgap energy. As the number of half-cycles increases, high harmonics of the fundamental laser frequency emerge inside the bulk. Noticeably, the amplitude of the photo-current gradually decreases as the polarization angle rotates from $\theta = 0°$ to $\theta = 90°$, reflecting the change of the photo-excitation mechanism (cf. also Fig.12).

Thus depending on $\theta$, photo-excitation of electron-hole pairs and hence HHG is sensitive to the characteristics of the anisotropic multiple band structure of the solid. The discussion above was focused on the dynamics of laser-driven Bloch electrons at the Brillouin zone center with $\vec{k} = (0,0,0)$, however the photo-induced electric current associated with HHG is composed of microscopic currents from all $\vec{k}$-points in the Brillouin zone. In this work, we sample the Brillouin zone by a small number of points equidistantly spaced along the laser polarization direction with $\vec{k} = k\vec{e}$. The polarization angle-dependent final probability for electron–hole pair excitation as function of the length $k$ of the crystal momentum is shown in Fig.14 (a-c). The distribution is structured by sharp peaks associated with above-threshold excitation of electron-hole pairs. As discussed in Ref. [27], when the propagation phase of laser-driven electron-hole pair for one-cycle of the laser oscillations $S = nh$ is quantized with integer numbers multiple of the Planck's constant, resonant-like excitation of electrons into the conduction band occurs via constructive (Stuckelberg) interference of consecutive Landau-Zener transitions. For instance, in the example of the two coupled bands (v6, c1) shown in Fig.14 (a) ($\theta = 0°$), the two peaks at $k \approx \pm 0.05(2\pi/a)$ are a result of constructive interference of transitions with $S = 11h$. The increase of the action by one quantum, produces second pair of side peaks at $k \approx \pm 0.085(2\pi/a)$. As the polarization angle rotates in the (11-20) plane, the occupation of the conduction band from multiple valence bands becomes prominent. In Fig.14b ($\theta = 30°$), the filling of the conduction band occurs by the direct v6 → c1 transition and by the indirect transition pathway v4→ (v6, v7, v8) → c1, thus the peaks in the momentum distribution exhibit composite structure associated with different inter-band transitions. In Fig.14 (b) $\theta = 90°$, new pair of side peaks with $k \approx \pm 0.01(2\pi/a)$ emerge as a result of the v7→c1 inter-band transition. Note also the overall decrease of the probability for photo-excitation as the angle rotates from 0 to 90 degrees. The associated orientation-dependent HHG spectra can be obtained in very good approximation from the coherent sum of a small number of microscopic inter-band currents $\sum_r J(\vec{k}_r, t)$ of electrons having canonical crystal momenta $\vec{k}_r = k_r \vec{e}$ determined from the sharp peaks in the photoelectron yield. Thus, the angular modulation of the photo-electron yield translates into angular modulation of the orientation-dependent HHG yield.

The orientation-dependent high-harmonic spectra are shown in Fig.15-16 (a, b). In the (11-20) plane, Fig. 15 (a-b), the angular modulation of the harmonic intensity has two-fold symmetry, even-order harmonics above the band edge appear when the laser polarization direction is along the optical axis ( $0°$ or $180°$), for

$\theta = 90°$, only odd order harmonics appear. The harmonic spectra in the (0001) plane, shown in Fig.16 (a, b) extend beyond the band edge and display only odd harmonic orders, their intensity has angular modulation which clearly exhibits the six-fold rotational symmetry of the hexagonal lattice. Importantly, the orientation-dependent HHG yield depends also sensitively on the laser wavelength (cf. 3 µm versus 3.2 µm laser wavelength in Figs.15-16. For instance for the 3.2 µm wavelength, the harmonic orders 8-11 in the (11-20) plane fuse, creating a broad high-intensity structure in the frequency domain, which closely resembles the experimentally observed broadening of the harmonics from the 200 µm thick sample.

**Conclusion**

In conclusion, we generate high-order harmonics during irradiation of (0001) cut and (1120) cut ZnO samples of different thickness during irradiation with intense femtosecond mid-infrared laser pulses of wavelength 3.2 µm, pulse duration 45 fs and varying laser intensity from a high-power laser system with a repetition rate of 100 kHz. Laser intensities are chosen for which the intensities of the generated high harmonics are well visible for a wide range of harmonic orders extending to the XUV, and there no damage of the samples. Our main observation is that the harmonic spectra depend sensitively on the orientation of the crystal with respect to the laser polarization direction, with odd harmonics exhibiting periodicity of 45° in both (0001) plane and (1120) plane and even harmonics exhibiting periodicity of 90° in the (1120) plane.

The theoretical model is in good qualitative agreement with the experiment, in particular the strong dependence of the harmonic intensity on the polarization angle. This modulation arises from the intrinsic anisotropy of the ZnO band structure, which supports multiple coupled valence bands undergoing tunnel transitions into the conduction band.

Within the explored ranges of laser intensity and wavelength, the primary difference between theory and experiment lies in the observed four-fold symmetry in the measured signal, whereas the model yields the expected three-fold modulation dictated by the hexagonal lattice symmetry. This difference may originate from sensitive dependencies on the exact laser wavelength and peak intensity, suggesting that fine-tuning these parameters could reconcile the two.

Acknowledgements: This material is based upon work supported by the Air Force Office of Scientific Research under award number FA8655-24-1-7014. KP-06-COST/26 and ELIUPM4-92 are acknowledged.

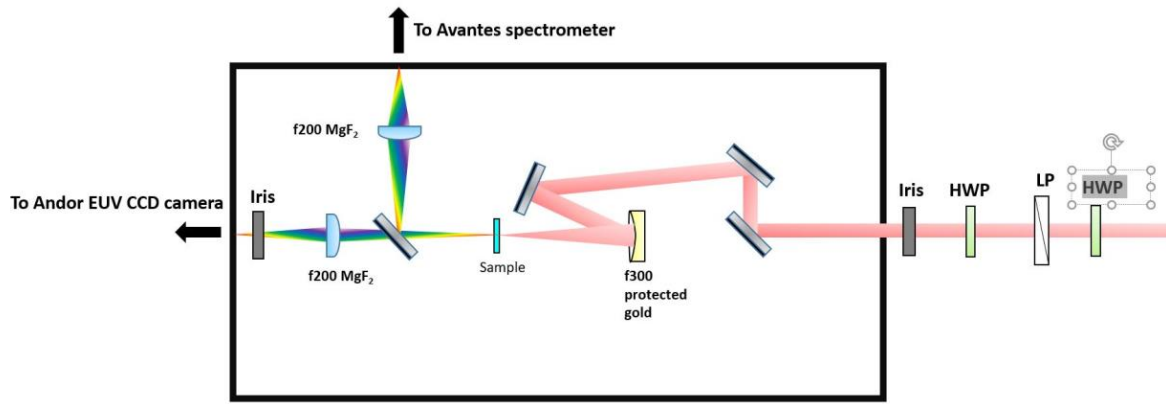

*Fig 1*

*Experimental setup*

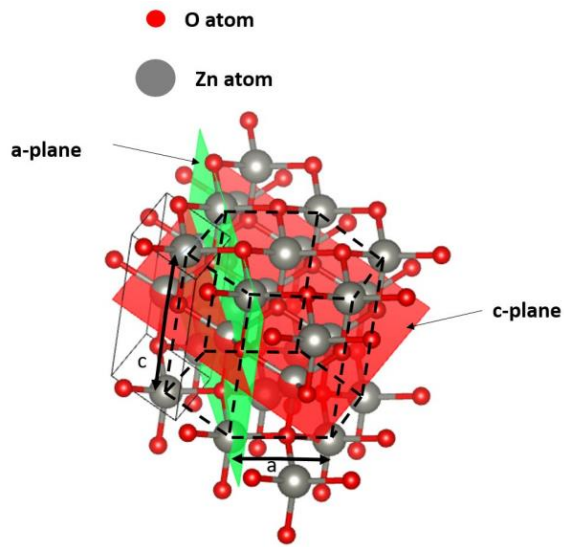

*Fig 2*

*Crystal structure ZnO*

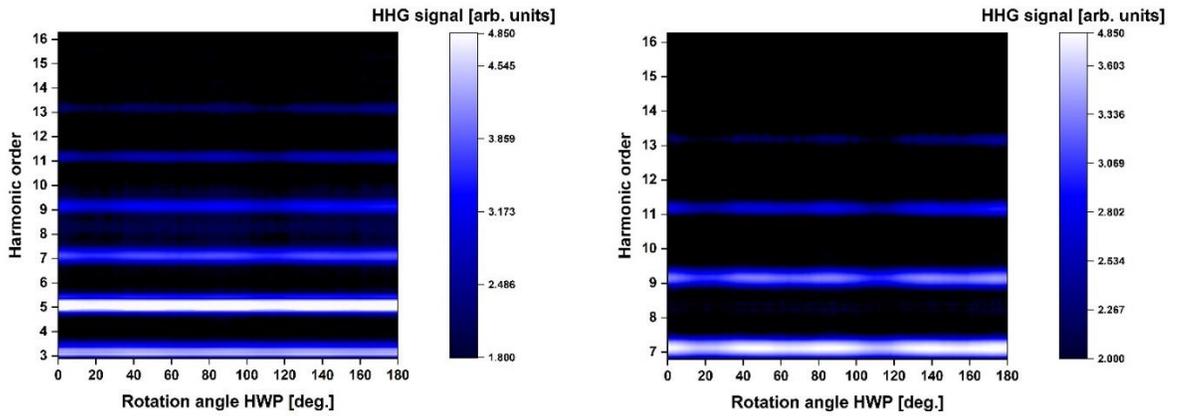

*Fig 3*

*High harmonic generation from c- cut (plane 0001) ZnO with thickness 50 µm in transmission geometry. The laser polarization in the c-plane and is varied by rotating half wave plate placed in the laser beam path. Main results – odd harmonics only; strong intensity modulation of above band gap harmonics. Laser intensity I = 1 TW/cm$^2$. Left panel – all harmonics. Right panel – only above band gap harmonics.*

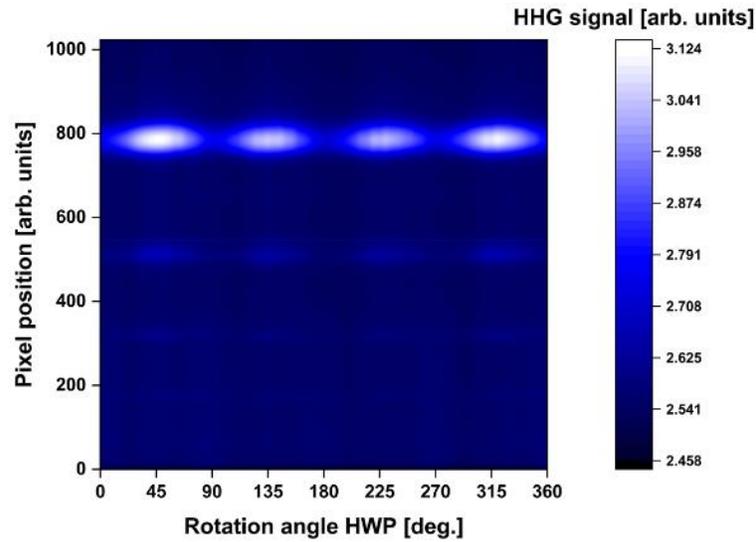

*Fig4*

*High harmonic generation from c- cut (plane 0001) ZnO with thickness 50 µm in transmission geometry. The laser polarization in the c-plane and is varied by rotating half wave plate (HWP) placed in the laser beam path, 13th, 15$^{th}$, 17$^{th}$ and 19$^{th}$ HO are observed.*

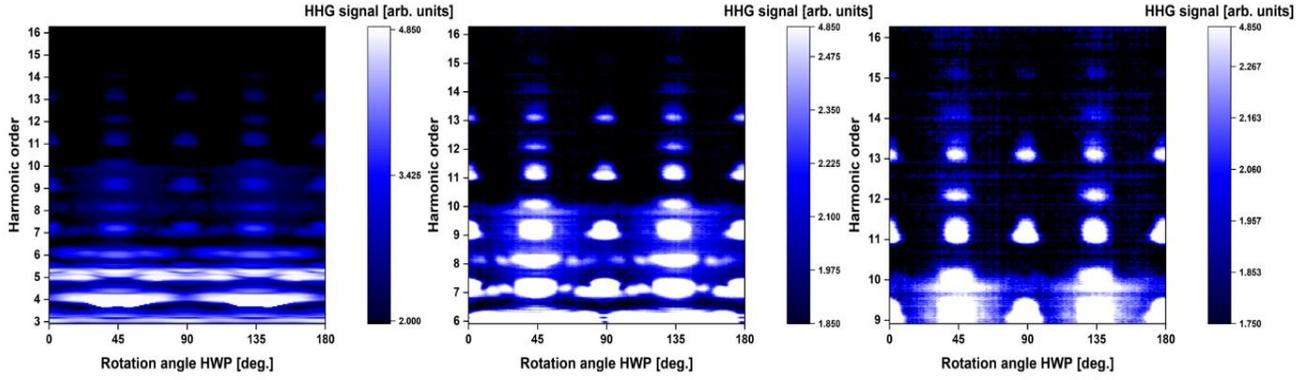

*Fig 5*

*High harmonic generation from a - cut (plane 1120) ZnO with thickness 50 µm in transmission geometry. The laser polarization in the a-plane and is varied by rotating half wave plate (HWP) placed in the laser beam path.*
*Generation of odd and even order harmonics, pronounced modulation of harmonic intensity with laser polarization direction.*

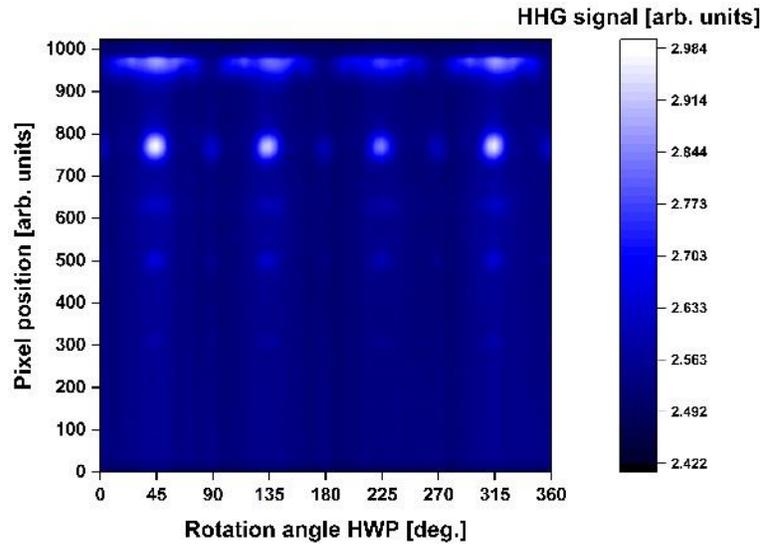

*Fig 6*

*High harmonic generation from a - cut (plane 1120) ZnO with thickness 50 µm in transmission geometry. The laser polarization in the a-plane and is varied by rotating half wave plate (HWP) placed in the laser beam path. $13^{th}$, $14^{th}$, $15^{th}$, $16^{th}$, $17^{th}$, $18^{th}$ and $19^{th}$ HO are generated.*

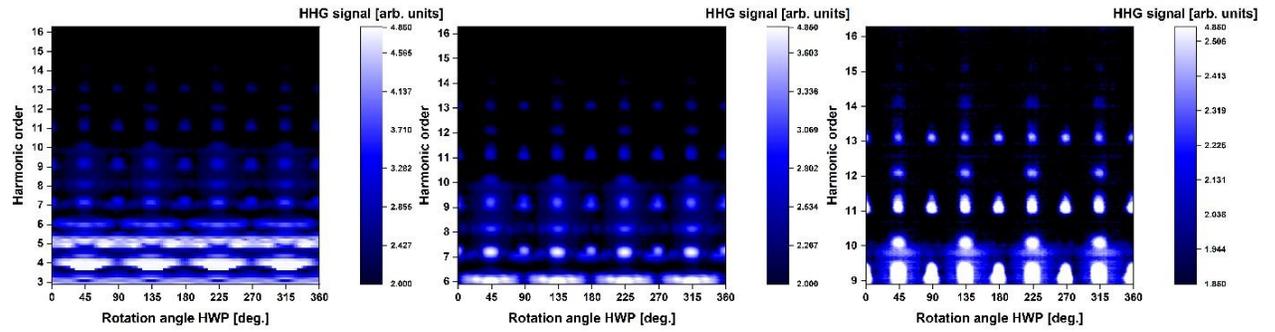

*Fig 7*

*High harmonic generation from a - cut (plane 1120) ZnO with thickness 50 μm in transmission geometry The laser polarization in the a-plane and is varied by rotating half wave plate (HWP) placed in the laser beam path. ZnO is a polar crystal.*

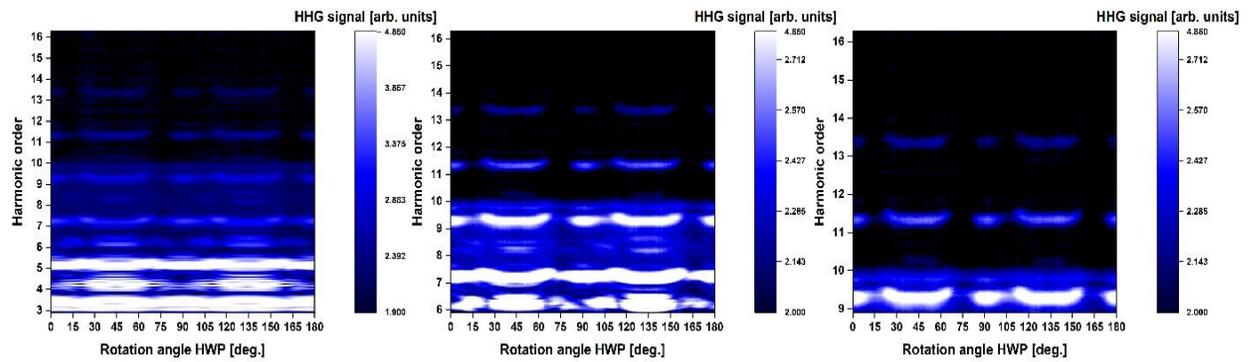

*Fig 8*

*High harmonic generation from a - cut (plane 1120) ZnO with thickness 200 μm in transmission geometry. The laser polarization in the a-plane and is varied by rotating half wave plate (HWP) placed in the laser beam path. I = 1.2 TW/cm$^2$*

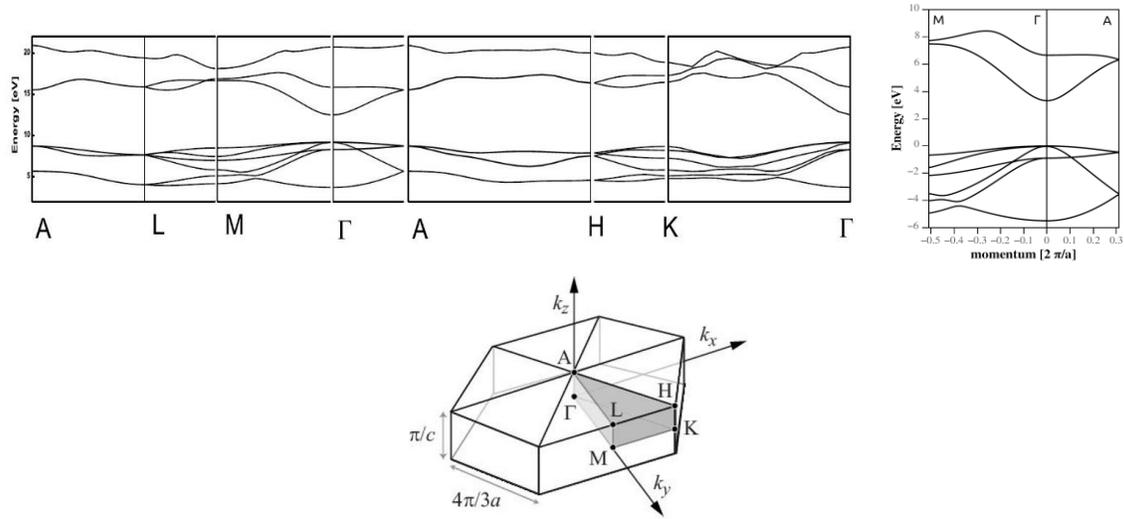

*Fig 9*

*Top - static band structure of zinc oxide along the Δ (Γ-A) and Σ (Γ-M) lines in the Brillouin zone. Bottom - Brillouin zone of ZnO: circles show high symmetry points. The shaded area is the primitive unit cell of the Brillouin zone. The lattice constants a and c are 3.25 and 5.21 (Å) respectively. The crystal momentum is measured in units 2π/a.*

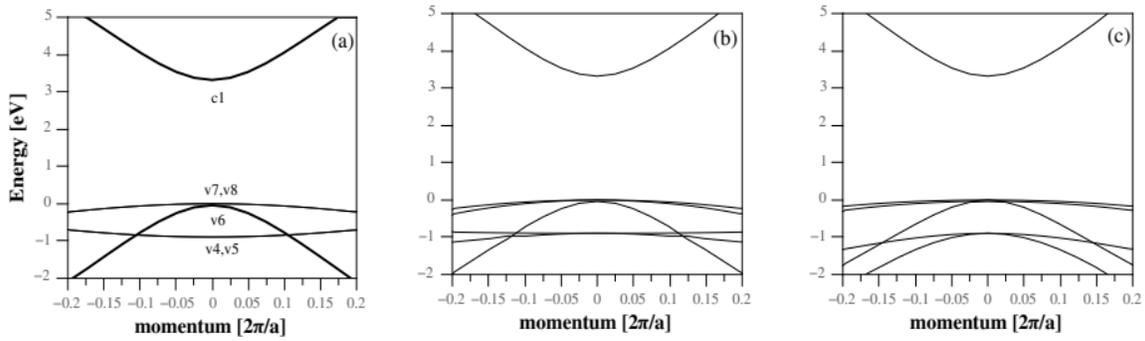

*Fig.10*

*Variation of the static band structure by changing the polarization angle. The lowest in energy conduction band, together with five valence bands of ZnO are shown. The polarization angle is rotated in the (11-20) plane of the crystal. (a) $\theta = 0°$, (b) $\theta = 30°$ and (c) $\theta = 90°$*

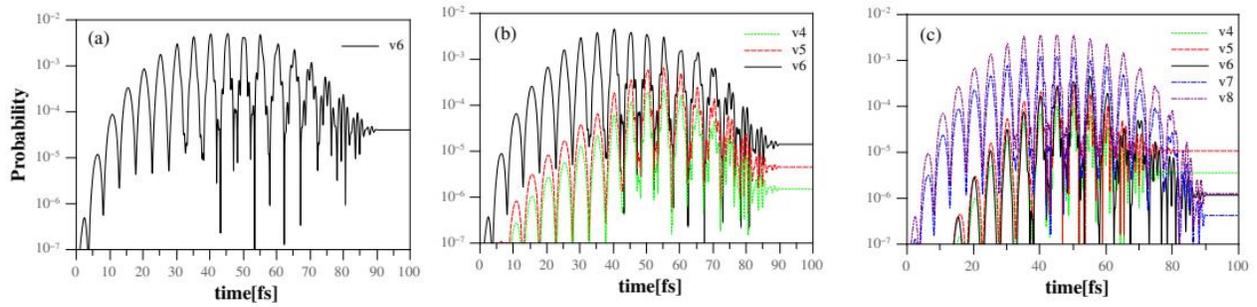

*Fig. 11*

*Probability for photo-excitation of electrons in the Γ-bands of ZnO subjected to intense 9-cycle MIR laser pulse having laser wavelength 3 μm with peak laser intensity 2 TW/cm². In Fig (a-c), the polarization angle $\theta$ is varied in the (11-20) plane of the crystal: Fig.11 (a) $\theta = 0°$, in Fig. (b) $\theta = 30°$ and in Fig. (c) $\theta = 90°$. The different labels designate partial contribution of each valence band to the electron yield in the conduction band.*

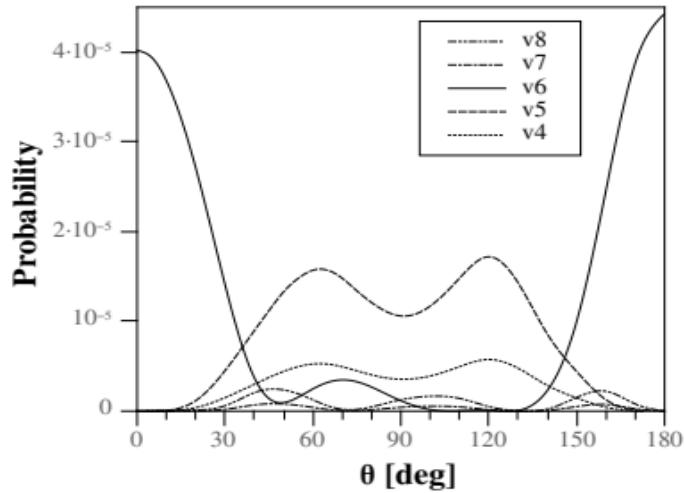

*Fig 12*

*Final probability for electron-hole pair excitation from the Γ-bands of ZnO crystal subjected to linearly polarized intense 9-cycle MIR laser pulse having laser wavelength 3 μm with peak laser intensity 2 TW/cm². The polarization angle $\theta$ is varied in the (11-20) plane of the crystal. The different labels designate partial contribution of each valence band to the electron yield in the conduction band.*

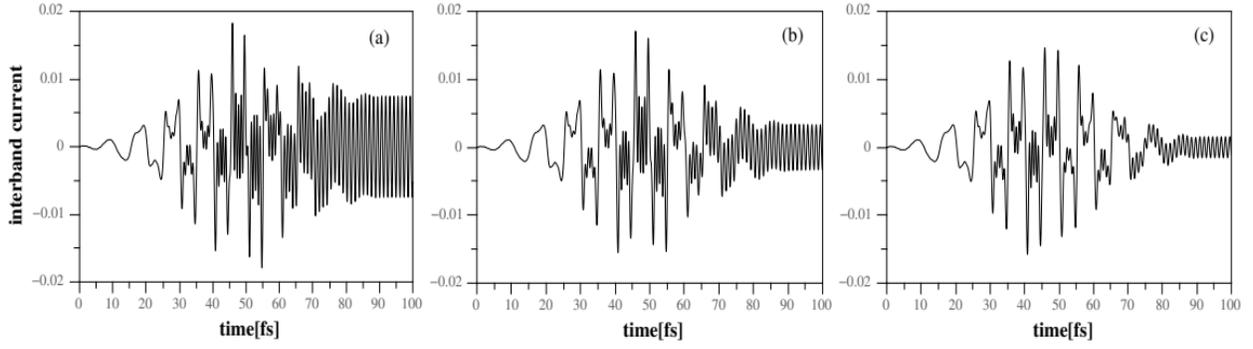

*Fig13.*

*Time-dependent inter- band current of electrons in the $\Gamma$ -bands of ZnO subjected to intense 9-cycle MIR laser pulse, having laser wavelength 3 μm with peak laser intensity 2 TW/cm$^2$ . In Fig. 13 (a-c), the polarization angle $\theta$ is varied in the (11-20) plane of the crystal: Fig(a) $\theta = 0°$ , in Fig. (b) $\theta = 30°$ and in Fig. (c) $\theta = 90°$ .*

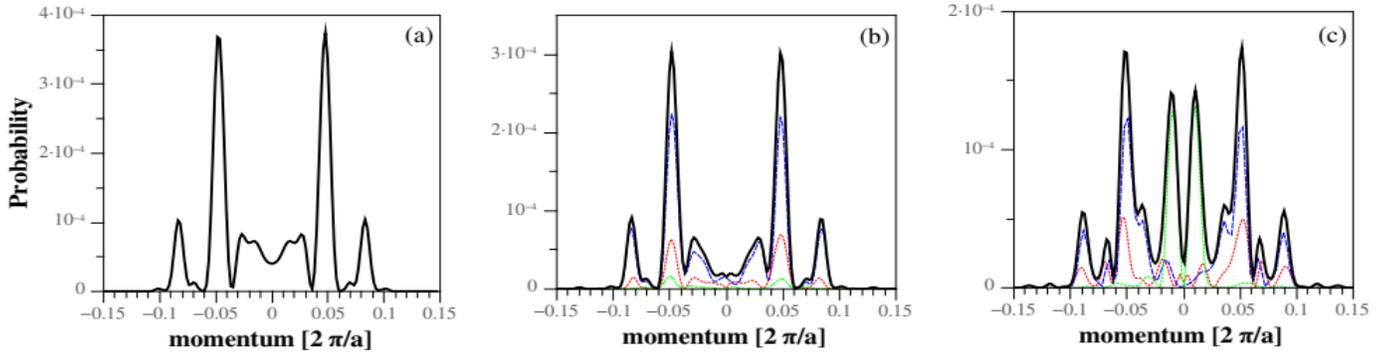

*Fig. 14*

*Crystal momentum dependent photo-electron yield. The laser polarization direction is varied in the (11-20) plane. The laser intensity is I = 2 TW/cm$^2$, the laser wavelength is 3 microns, the pulse duration is 90 fs. The labels designate partial contributions of each valence band. (a) $\theta = 0°$ , (b) $\theta = 30°$ and (c) $\theta = 90°$ The dashed (blue) line in (b, c) corresponds to the v6-c1 interband transition, the dotted (red) line (v4-c1) and dashed-dotted (green) line ( v7-c1) . In Fig. 14 (a), only (v6, c1) transition is allowed by selection rules.*

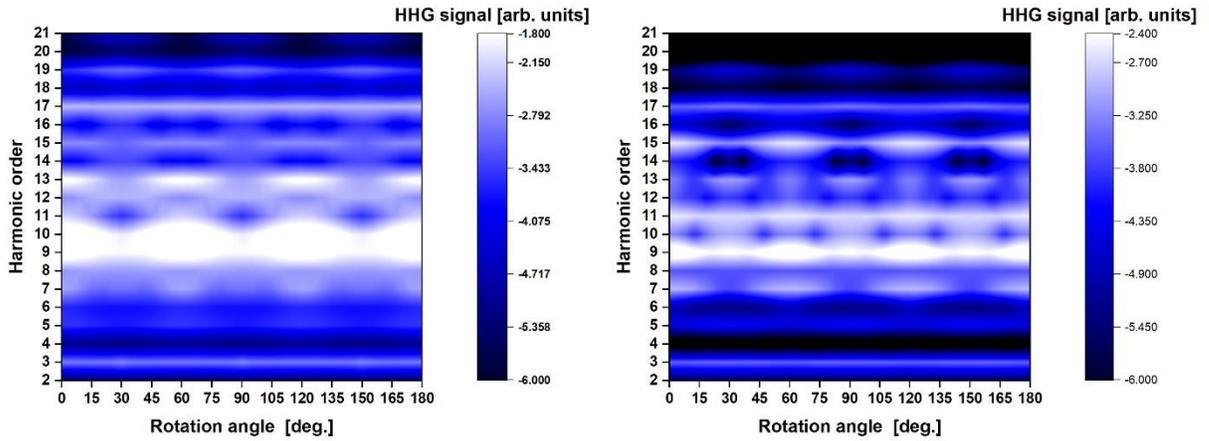

*Fig 15*

*Theoretical calculation of the generated high harmonics from c- cut (plane 0001) ZnO for laser intensity. I = 2 TW/cm$^2$. Left panel – wavelength λ=3.2 μm, right panel – wavelength λ=3 μm. The other laser parameters are given in the text.*

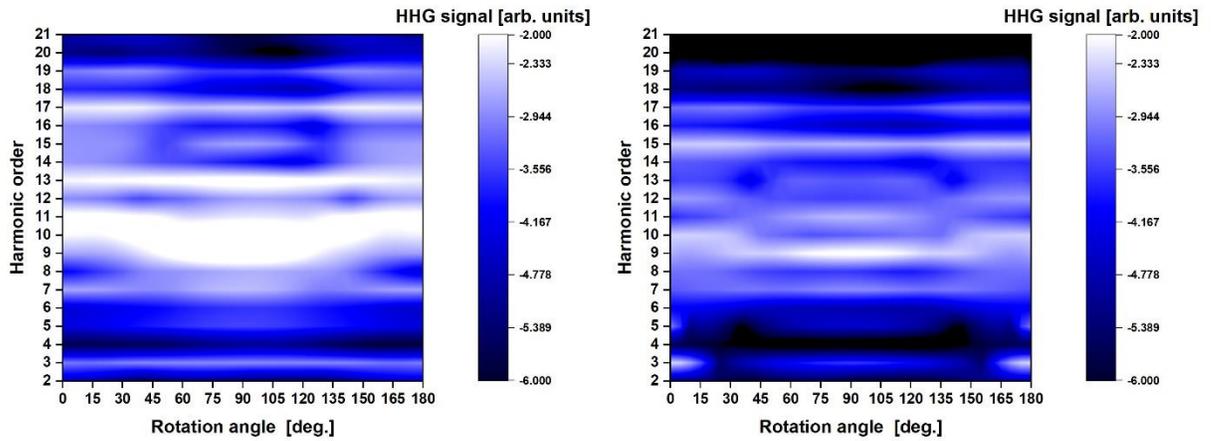

*Fig 16*

*Theoretical calculation of the generated high harmonics from a - cut (plane 1120) ZnO for laser intensity. I = 2 TW/cm$^2$. Left panel – wavelength λ=3.2 μm, right panel – wavelength λ=3 μm. The other laser parameters are given in the text.*

**References:**

[1] E. Goulielmakis, M. Schultze, M. Hofstetter, V. S. Yakovlev, J. Gagnon, M. Uiberacker, A. L. Aquila, E. Gullikson, D. T. Attwood, R. Kienberger et al., "Single-cycle nonlinear optics," Science 320, 1614 (2008).

[2] S. Haessler, J. Caillat, W. Boutu, C. Giovanetti-Teixeira, T. Ruchon, T. Auguste, Z. Diveki, P. Breger, A. Maquet, B. Carré et al., "Attosecond imaging of molecular electronic wavepackets," Nat. Phys. 6, 200, (2010)

[3] P. Peng, C. Marceau, and D. M. Villeneuve, "Attosecond imaging of molecules using high harmonic spectroscopy," Nat. Rev. Mater. 1, 144, (2019)

[4] G. Sansone, "Looking into strong-field dynamics," Nat. Photonics 14, 131, (2020)

[5] S. Y. Kruchinin, F. Krausz, and V. S. Yakovlev, "Colloquium: Strong-field phenomena in periodic systems," Rev. Mod. Phys. 90(2), 021002 (2018)

[6] J. Weisshaupt, V. Juvé, M. Holtz, S. Ku, M. Woerner, T. Elsaesser, S. Ališauskas, A. Pugžlys, and A. Baltuška, "High-brightness table-top hard X-ray source driven by sub-100-femtosecond mid-infrared pulses," Nat. Photonics 8(12), 927–930 (2014)

[7] A. S. Johnson, D. R. Austin, D. A. Wood, C. Brahms, A. Gregory, K. B. Holzner, S. Jarosch, E. W. Larsen, S. Parker, C. S. Strüber, P. Ye, J. W. G. Tisch, and J. P. Marangos, "High-flux soft x-ray harmonic generation from ionization-shaped few-cycle laser pulses," Sci. Adv. 4(5), eaar3761 (2018)

[8] B. Wolter, M. G. Pullen, A. T. Le, M. Baudisch, K. Doblhoff-Dier, A. Senftleben, M. Hemmer, C. D. Schröter, J. Ullrich, T. Pfeifer, R. Moshammer, S. Gräfe, O. Vendrell, C. D. Lin, and J. Biegert, "Ultrafast electron diffraction imaging of bond breaking in di-ionized acetylene," Science 354(6310), 308–312 (2016)

[9] V. Shumakova, P. Malevich, S. Ališauskas, A. Voronin, A. M. Zheltikov, D. Faccio, D. Kartashov, A. Baltuška, and A. Pugžlys, "Multi-millijoule few-cycle mid-infrared pulses through nonlinear self-compression in bulk," Nat. Commun. 7(1), 12877 (2016)

[10] T. P. Butler, D. Gerz, C. Hofer, J. Xu, C. Gaida, T. Heuermann, M. Gebhardt, L. Vamos, W. Schweinberger, J. A. Gessner, T. Siefke, M. Heusinger, U. Zeitner, A. Apolonski, N. Karpowicz, J. Limpert, F. Krausz, and I. Pupeza, "Watt-scale 50-MHz source of single-cycle waveform-stable pulses in the molecular fingerprint region," Opt. Lett. 44(7), 1730–1733 (2019)

[11] S. Gholam-Mirzaei, J. E. Beetar, A. Chacón, and M. Chini, "High-harmonic generation in ZnO driven by self-compressed mid-infrared pulses," J. Opt. Soc. Am. B 35(4), A27–A31 (2018)

[12] S. Ghimire, A. D. DiChiara, E. Sistrunk, et al. "Redshift in the optical absorption of ZnO single crystals in the presence of an intense midinfrared laser field". Phys. Rev. Lett. 107, 167407, (2011)

[13] S. Ghimire, A. D. DiChiara, E. Sistrunk, P. Agostini, L. F. DiMauro and D. A. Reis, Nat. Phys. 7, 138 (2011)

[14] S. Gholam-Mirzaei, J. Beetar, and M. Chini, Appl. Phys. Lett. 110, 061101 (2017)

[15] O. Schubert, M. Hohenleutner, F. Langer, B. Urbanek, C. Lange, U. Huttner, D. Golde, T. Meier, M. Kira, S. W. Koch and R. Huber Nat. Photon 8, 119, (2014)

[16] G. Vampa, T. J. Hammond, N. Thire, B. E. Schmidt, F. Legare, C. R. McDonald, T. Brabec, D. D. Klug, and P. B. Corkum, All-Optical Reconstruction of Crystal Band Structure, Phys. Rev. Lett. 115, 193603 (2015)

[17] G. Vampa, T. J. Hammond, N. Thire, B. E. Schmidt, F. Legare, C. R. McDonald, T. Brabec and P. B. Corkum, Nature 522, 462 (2015)

[18] Jiang, S. et al. "Crystal symmetry and polarization of high-order harmonics in ZnO", Journal of Physics B 52, 225601(2019)

[19] Z. Wang, H. Park, Y. H. Lai, J. Xu, C. I. Blaga, F. Yang, P. Agostini, and L. F. DiMauro, Nat. Commun. 8, 1686 (2017)

[20] H. Liu, Y. Li, Y. S. You, S. Ghimire, T. F. Heinz, and D. A. Reis, High-harmonic generation from an atomically thin semiconductor, Nat. Phys. 13, 262 (2017)

[21] F. Langer, M. Hohenleutner, U. Huttner, S. W. Koch, M. Kira, and R. Huber, Symmetry-controlled temporal structureof high-harmonic carrier fields from a bulk crystal, Nat. Photonics 11, 227 (2017)

[22] T. T. Luu, M. Garg, S. Y. Kruchinin, A. Moulet, M. T. Hassan, and E. Goulielmakis, Extreme ultraviolet high-harmonic spectroscopy of solids, Nature (London) 521, 498 (2015)

[23] L. Li, P. Lan, L. He, W. Cao, Q. Zhang, and P. Lu, Phys. Rev. Lett. 124, 157403 (2020)

[24] T. T. Luu and H. J. Wörner, Nat. Commun. 9, 916 (2018)

[25] N. Thiré, R. Maksimenka, B. Kiss, C. Ferchaud, G. Gitzinger, T. Pinoteau, H. Jousselin, S. Jarosch, P. Bizouard, V. Di Pietro, E. Cormier, K. Osvay, and N. Forget, "Highly stable, 15 W, few-cycle, 65 mrad CEP-noise mid-IR OPCPA for statistical physics," Opt. Express 26(21), 26907–26915 (2018)

[26] F. Médard, "Conception et spectroscopie de microcavités à base de ZnO en régime de couplage fort pour l'obtention d'un laser à polaritons", PhD dissertation, UNIVERSITÉ BLAISE PASCAL – CLERMONT-FERRAND II, 2010

[27] Obreshkov, B., Apostolova, T. High-harmonic generation in zinc oxide subjected to intense mid-infrared femtosecond laser pulse. *Appl. Phys. B* **131**, 158 (2025), https://doi.org/10.1007/s00340-025-08506-y

[28] M. Cohen, T. K. Bergstresser, Phys. Rev. 141, 789 (1966).